# MIND: A Noise-Adaptive Denoising Framework for Medical Images Integrating Multi-Scale Transformer

Tao Tang
School of Computer Science and Engineering
University of Electronic Science and Technology of China
Chengdu, China
tangtao_cs@std.uestc.edu.cn

Chengxu Yang*
School of Computer Science and Engineering
University of Electronic Science and Technology of China
Chengdu, China
yangcx@uestc.edu.cn

*Abstract*—The core role of medical images in disease diagnosis makes their quality directly affect the accuracy of clinical judgment. However, due to factors such as low-dose scanning, equipment limitations and imaging artifacts, medical images are often accompanied by non-uniform noise interference, which seriously affects structure recognition and lesion detection. This paper proposes a medical image adaptive denoising model (MIND) that integrates multi-scale convolutional and Transformer architecture, introduces a noise level estimator (NLE) and a noise adaptive attention module (NAAB), and realizes channel-spatial attention regulation and cross-modal feature fusion driven by noise perception. Systematic testing is carried out on multimodal public datasets. Experiments show that this method significantly outperforms the comparative methods in image quality indicators such as PSNR, SSIM, and LPIPS, and improves the F1 score and ROC-AUC in downstream diagnostic tasks, showing strong practical value and promotional potential. The model has outstanding benefits in structural recovery, diagnostic sensitivity, and cross-modal robustness, and provides an effective solution for medical image enhancement and AI-assisted diagnosis and treatment.

*Keywords- Signal Processing; Computer Vision; Bioinformatics; Artificial Intelligence; Information Fusion*

## I. Introduction

Medical images are vital for diagnosis but often suffer from noise due to factors like low-dose scanning, which reduces diagnostic reliability by interfering with lesion identification. Traditional filtering and static deep learning methods have improved denoising but possess clear limitations. They struggle to handle different noise types and intensities, and perform poorly in structure preservation and cross-modal generalization, making it difficult for them to meet clinical needs. To solve these problems, this paper builds an adaptive denoising model for medical images using an attention mechanism. By introducing a noise level perception mechanism and a multi-scale Transformer, it achieves dynamic, adaptive processing in noisy environments. The model optimizes image quality indicators like PSNR, SSIM, and LPIPS, and also improves the accuracy and robustness of downstream diagnostic tasks, showing broad clinical and engineering application value.

## II. Related Work

### 2.1 Overview of Medical Image Denoising Methods

Medical image denoising has evolved from traditional filtering to deep learning. Zhang et al. [1] proposed a CT denoising model using a U-Net with multiple attention mechanisms, which improves structural restoration and enhances PSNR and SSIM by focusing on lesion areas. Kulathilake et al. [2] reviewed adaptive denoising methods, summarizing various strategies and noting current challenges in generalization, modality compatibility, and noise modeling accuracy. Huang et al. [3] built a CNN integrating anatomical priors and attention for low-dose CT, achieving efficient noise suppression and demonstrating the value of structural priors. These studies have advanced the field from static filtering to adaptive, structure-aware, intelligent denoising.

### 2.2 Attention Mechanism and Visual Transformer

Following the success of the Transformer architecture in computer vision, the attention mechanism is now widely used in medical image denoising to model structural and contextual features (Figure 1). Jain et al. [4] proposed a pyramid denoising network with an optimal attention block to improve edge preservation and semantic consistency. Lee et al. [5] designed a multi-scale self-attention network that is well-suited for low-dose image processing scenarios. Chen et al. [6] integrated attention into image segmentation to improve model robustness in high-noise images. Xiong et al. [7] proposed the Ina-Net architecture with a noise-adaptive gated attention mechanism, allowing the network to dynamically adjust its strategy based on regional noise characteristics. These studies show that visual Transformers and attention offer new approaches for modeling medical images in complex noise conditions.

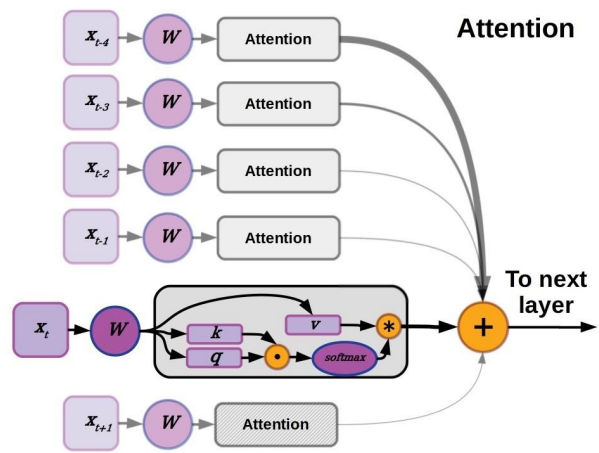

Figure 1. Transformer architecture diagram

## 2.3 Joint Learning of Super-Resolution and Denoising

Joint modeling of image super-resolution and denoising is an important direction for improving image reconstruction quality. Sharif et al. [8] proposed a dynamic residual attention network using residual blocks and attention to coordinate image resolution restoration and noise elimination. Song et al. [9] built APNet with an adaptive projection mechanism to restore texture while reducing noise, thus improving perceptual quality. Gonçalves et al. [10] noted that multi-task learning will become a mainstream design, and that attention mechanisms can effectively coordinate feature sharing in joint super-resolution and denoising tasks. This prior research provides theoretical support for the modules used in this study.

## III. METHODS

### 3.1 Overall Framework

The overall structure of the medical image adaptive denoising model (MIND), shown in Figure 2, has five modules: a multi-scale residual pyramid encoder-decoder, a Transformer cascade module, a cross-modal feature fusion module, a noise level estimator (NLE), and a noise adaptive attention block (NAAB). First, the original noisy image is processed by the multi-scale encoder-decoder to extract features. These features are then enhanced by a Transformer Cascade Module to model long-distance dependencies. In parallel, the Cross-Modal Fusion module extracts and aligns features from the original image, a preliminary denoised image, and its gradient map. The fused features are then sent to the NLE to generate noise perception parameters, $\gamma$ and $\beta$. The NAAB module uses these parameters to dynamically adjust its channel and spatial attention paths, allowing it to adapt to noise intensity in different regions. Finally, the model outputs a high-fidelity denoised image, optimizing for both structural preservation and perceptual quality through an end-to-end loss function.

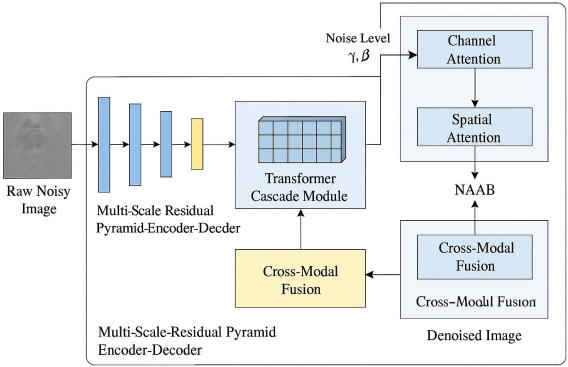

Figure 2. Overall structure of the medical image adaptive denoising model

### 3.2 Noise-Level Estimator (NLE)

Because noise in medical images is not uniformly distributed and varies with modality, imaging site, and dose, this study designed an unsupervised noise level estimator (NLE). To improve the model's regional adaptability, the NLE dynamically estimates noise intensity in different areas of the input image. The image degradation model is considered as:

$$\mathbf{Y} = \mathbf{X} + \mathbf{N} \quad (1)$$

In this model, $\mathbf{Y}$ represents the noisy image, $\mathbf{X}$ is the ideal noise-free image, and $\mathbf{N}$ is the additive noise term. Assuming the noise $\mathbf{N}$ has a zero mean and a position-dependent standard deviation $\sigma(i,j)$, the noise level can be estimated using local statistics of the gradient residual via a sliding window method. The local residual estimate is defined as:

$$\sigma(i,j) = \sqrt{\mathbb{E}\left[(\nabla Y(i,j))^2\right] - (\nabla \hat{X}(i,j))^2} \quad (2)$$

Here, $\nabla Y(i,j)$ and $\nabla \hat{X}(i,j)$ are the gradient responses of the input image and the preliminary estimated image, respectively. They are usually calculated using the Sobel or Scharr operator. This estimation method is label-independent and can be iteratively optimized during training. The estimate $\hat{X}$ is obtained from the output of the multi-scale residual encoder-decoder branch before the Transformer cascade, providing a coarse denoised image for noise estimation.

Finally, the NLE module outputs a parameterized noise map $\sigma(i,j)$ and generates two sets of control parameters through two sets of learnable transformation functions $\gamma$, $\beta$:

$$\gamma = f_\gamma(\sigma), \quad \beta = f_\beta(\sigma) \quad (3)$$

The above function is implemented through a shallow CNN to achieve the mapping from the noise estimation map to the attention modulation factor for use by the subsequent NAAB.

### 3.3 Noise Level Adaptive Attention Block (NAAB)

The NAAB aims to use the noise perception parameters output by NLE to achieve spatial and channel attention modulation to adapt to the noise level in varying regions and boost the selectivity and accuracy of denoising. Its input is the encoded feature map $\mathbf{F} \in \mathbb{R}^{C \times H \times W}$, where $C, H, W$ are the number of channels, height and width respectively.

The NLE outputs $\gamma$ and $\beta$ are then used for normalization:

$$\mathbf{F}' = \gamma \cdot \mathbf{F} + \beta \quad (4)$$

Here, $\gamma, \beta \in \mathbb{R}^{C \times 1 \times 1}$, a broadcast mechanism is applied to each channel to recalibrate channels based on the noise level.

Then, the channel attention branch calculates the global average pooling vector $\mathbf{z} \in \mathbb{R}^C$:

$$z_c = \frac{1}{H \times W} \sum_{i=1}^{H} \sum_{j=1}^{W} F'_{c,i,j} \quad (5)$$

And through the fully connected layer and nonlinear transformation, the channel weight vector is obtained $\alpha \in \mathbb{R}^C$:

$$\alpha = Sigmoid\left(W_2 \cdot ReLU\left(W_1 \cdot \mathbf{z}\right)\right) \quad (6)$$

where $W_1 \in \mathbb{R}^{C/r \times C}, W_2 \in \mathbb{R}^{C \times C/r}$, $r$ is the compression ratio.

The parallel spatial attention branch uses convolution to capture local responses and generate a weight map $\mathbf{A}_{spatial} \in \mathbb{R}^{1 \times H \times W}$:

$$\mathbf{A}_{spatial} = Sigmoid\left(Conv_{7\times 7}\left(AvgPool(\mathbf{F}) \| MaxPool(\mathbf{F})\right)\right) \quad (7)$$

Finally, the fused output is:

$$\mathbf{F}_{att} = \mathbf{F}' \cdot \alpha \otimes \mathbf{A}_{spatial} \quad (8)$$

In this equation, "·" denotes element-wise multiplication with channel-wise broadcasting, and "⊗" denotes element-wise multiplication with spatial broadcasting. This mechanism uses dual attention to suppress noise-dominated areas while preserving structural details across noise levels, improving perceptual selectivity and robustness.

### 3.4 Cross-Modal Feature Interaction and Information Distillation

In Figure 3, the cross-modal feature interaction module integrates three inputs: the original noisy image $\mathbf{1}_n$, the preliminary denoised image $\mathbf{1}_d$, and the gradient map $\mathbf{G}$, which capture the overall noise, rough structure, and edge detail information respectively. The three are sequentially converted into a unified vector sequence through the linear projection layer, and then input into the self-attention module in the Transformer architecture to achieve cross-modal deep information distillation.

Suppose the original input is encoded as:

$$\mathbf{F}'_n = f_{enc}(\mathbf{1}_n), \quad \mathbf{F}'_d = f_{enc}(\mathbf{1}_d), \quad \mathbf{F}'_g = f_{enc}(\mathbf{G}) \quad (9)$$

Among them, $f_{enc}(\cdot)$ represents the linear projection function, the output features are $\mathbb{R}^{C \times H \times W}$, and the three modes are connected in series to form a sequence representation:

$$\mathbf{Z} = Concat\left(Flatten(\mathbf{F}_n), Flatten(\mathbf{F}_d), Flatten(\mathbf{F}_g)\right) \in \mathbb{R}^{L \times D} \quad (10)$$

$L = 3 \times H \times W$ is the sequence length, $D$ is the embedding dimension, $\mathbf{F}_n, \mathbf{F}_d, \mathbf{F}_g$ is the feature map encoded by 3 modes.

The Transformer's self-attention mechanism is below:

$$Attention(\mathbf{Q}, \mathbf{K}, \mathbf{V}) = Softmax\left(\frac{\mathbf{Q}\mathbf{K}^\top}{\sqrt{d_k}}\right)\mathbf{V} \quad (11)$$

where $\quad \mathbf{Q} = \mathbf{Z}W_Q, \quad \mathbf{K} = \mathbf{Z}W_K, \quad \mathbf{V} = \mathbf{Z}W_V$

where is $\mathbf{Z}$ the flattened and concatenated token sequence from all modalities. $W_Q, W_K, W_V$ represents the linear projection matrix of query, key and value respectively, $d_k$ is the dimension of the key vector. The final output cross-modal embedding is expressed as:

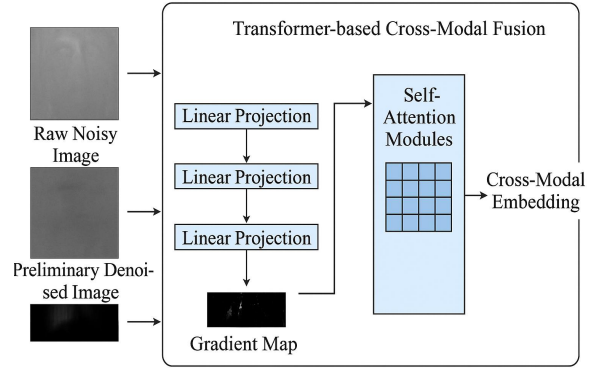

Figure 3. Schematic diagram of cross-modal feature interaction module

### 3.5 Loss Function Design

In order to balance pixel fidelity, structural similarity, edge preservation, and perceptual quality under different noise intensities and modalities, this paper designs a weighted combination loss function, and its total loss form is:

$$\mathcal{L}_{total} = \lambda_1(\sigma)\mathcal{L}_{MSE} + \lambda_2(\sigma)\mathcal{L}_{SSIM} + \lambda_3(\sigma)\mathcal{L}_{edge} + \lambda_4(\sigma)\mathcal{L}_{perc} + \lambda_5(\sigma)\mathcal{L}_{adv} \quad (12)$$

Among them, $\mathcal{L}_{MSE}$ represents the mean square error loss, which is used to maintain pixel accuracy:

$$\mathcal{L}_{MSE} = \frac{1}{N}\sum_{i=1}^{N} \| \hat{\mathbf{X}}_i - \mathbf{X}_i \|_2^2 \quad (13)$$

Among them, $\hat{\mathbf{X}}_i$ represents the pixel of the denoised image generated by the network $i$, $\mathbf{X}_i$ is the corresponding real noise-free image pixel, and $N$ is the total number of pixels in the image.

$\mathcal{L}_{SSIM}$ represents the structural similarity loss, which measures the consistency of local structure reconstruction:

$$\mathcal{L}_{edge} = \| \nabla\hat{\mathbf{X}} - \nabla\mathbf{X} \|_1 \quad (14)$$

Among them, $\nabla$ represents the edge extraction operator, which acts on the denoised image and the real image to preserve the boundary structure.

Perceptual loss uses a pre-trained model to extract feature space differences:

$$\mathcal{L}_{perc} = \| \phi(\hat{\mathbf{X}}) - \phi(\mathbf{X}) \|_2^2 \quad (15)$$

Among them, $\phi(\cdot)$ represents the intermediate feature extraction function in the perception network, reflecting the high-level semantic consistency of the image.

An optional adversarial loss is used to improve the realism of the image:

$$\mathcal{L}_{adv} = -\log D(\hat{\mathbf{X}}) \quad (16)$$

Among them, $D(\cdot)$ represents the confidence output of the discriminator that the generated image is a real image, which conforms with the objective function of the GAN framework.

In order to achieve adaptive adjustment of noise intensity, the following noise weighting strategy is introduced:

$$\lambda_i(\sigma) = \alpha_i \cdot \exp(-\beta_i \cdot \sigma) \quad (17)$$

Among them, $\lambda_i(\sigma)$ is $i$ the dynamic weight of the loss, $\sigma$ is the noise level estimated by the NLE module, $\alpha_i, \beta_i$ and is the preset weight scaling factor and decay rate, which is used to enhance MSE/edge control in high noise conditions and enhance perception and structural consistency in low noise conditions.

## IV. EXPERIMENTAL DESIGN

### 4.1 Dataset

We train and evaluate on public multimodal datasets: NIH ChestX-ray14 (>140,000 images of various lesions), BraTS 2023 (~1,250 T1/T1c/T2/FLAIR MRI sets for brain artifact and noise restoration), ACRIN-6698 breast MRI (385 chemotherapy-stage patients' DWI for cross-time noise consistency), and "Healthy-Total-Body CTs" (30 adults paired low-dose/high-dose 5mAs scans). Synthetic noise — Gaussian ($\sigma$=5-25%), Poisson, speckle (10-30%), and motion blur (5-15px)-is applied across all modalities, with aligned standard/low-dose PET/CT preserving realistic distributions. All images are cropped to 256×256 for batch training.

### 4.2 Implementation Details

Model training is performed on the NVIDIA A100 80GB GPU platform using the PyTorch 2.1.0 framework and CUDA version 11.8. The input image size is unified to 256×256 pixels, the batch size is set to 16, the optimizer is Adam, the initial learning rate is set to 1e-4, and the cosine decay strategy is used to dynamically decay to 1e-6 within 100 epochs.

The weight coefficient of the loss function is dynamically adjusted according to the noise level and is initially set to: $\alpha_{MSE} = 1.0$, $\alpha_{SSIM} = 0.8$, $\alpha_{edge} = 0.6$, $\alpha_{pere} = 0.4$, $\alpha_{adv} = 0.1$, where the exponential decay parameter $\beta_i$ is 0.15 and the noise intensity $\sigma$ is taken from the full-image mean output by the NLE module.

The multimodal input includes the original image, preliminary denoised image, and gradient image. Data enhancement includes random rotation, intensity perturbation, and shear perturbation. In order to improve the stability of cross-modal feature alignment, the Transformer module introduces residual connection and LayerNorm structure, and the fused features are updated through self-attention.

The following is the training implementation parameter configuration Table I:

TABLE I. TRAINING IMPLEMENTATION PARAMETER CONFIGURATION

| Parameter | Value/Setting |
|---|---|
| Input size | 256 × 256 |
| Batch size | 16 |
| Epochs | 100 |
| Optimizer | Adam |
| Initial learning rate | 1e-4 |
| LR schedule | Cosine Decay |
| Backbone | Multi-Scale Residual + Transformer |
| Loss Weights | MSE: 1.0, SSIM: 0.8, Edge: 0.6, Perc: 0.4, Adv: 0.1 |
| λ(σ) strategy | $\lambda = \alpha \cdot \exp(-\beta \cdot \sigma), \beta = 0.15$ |
| Hardware | NVIDIA A100 80GB × 2 |
| Framework | PyTorch 2.1.0 + CUDA 11.8 |

This configuration can achieve efficient convergence without relying on additional image labels and ensure the stability of the model in different modalities and noise scenarios.

### 4.3 Comparison Methods

This section compares MIND with three traditional methods (BM3D, NLM, Wiener) and four deep-learning denoisers (DnCNN, FFDNet, SwinIR, diffusion-based DDPM) on NIH ChestX-ray14 and BraTS 2023. All models are trained and normalized identically; results appear in Table II.

TABLE II. COMPARISON OF DENOISING METHODS

| Method | PSNR (dB) | SSIM | LPIPS ↓ | RMSE | F1 (X-ray) | ROC-AUC (MRI) |
|---|---|---|---|---|---|---|
| BM3D | 28.4 | 0.812 | 0.372 | 18.3 | 0.71 | 0.812 |
| NLM | 27.6 | 0.795 | 0.395 | 19.2 | 0.69 | 0.801 |
| Wiener | 26.9 | 0.781 | 0.401 | 20.1 | 0.68 | 0.785 |
| DnCNN | 30.2 | 0.845 | 0.292 | 16.7 | 0.75 | 0.84 |
| FFDNet | 30.5 | 0.854 | 0.285 | 16.4 | 0.76 | 0.851 |
| SwinIR | 31.8 | 0.879 | 0.223 | 14.8 | 0.8 | 0.875 |
| Diffusion (DDPM) | 32.1 | 0.887 | 0.207 | 14.3 | 0.82 | 0.883 |
| Ours (MIND) | 33.7 | 0.912 | 0.168 | 13.2 | 0.86 | 0.902 |

From quantitative indicators, traditional methods show limited PSNR and SSIM, and poor perceptual quality (LPIPS) and diagnostic F1 score. DnCNN and FFDNet improve mean-squared error metrics but fail at structural recovery. Experiments on public datasets confirm these trends. SwinIR and diffusion-based denoisers excel in LPIPS and AUC, especially diffusion preserving texture. The MIND model, leveraging noise-level estimation and cross-modal attention, yields the best results across multimodal datasets—PSNR 33.7 dB, SSIM 0.912, lesion F1 0.86—demonstrating robust detail consistency and diagnostic sensitivity.

*4.4 Ablation Experiments*

To further assess component contributions, we perform ablations by removing each key module—NAAB, NLE, multi-scale residual encoder, and cross-modal fusion—and repeating training/testing on the standard test set (Table III). Any removal significantly degrades performance: eliminating NAAB lowers PSNR to 31.6 dB, underscoring attention's role in noise-region perceptual regulation; removing NLE impairs noise modeling, with SSIM dropping from 0.912 to 0.889 and LPIPS rising from 0.168 to 0.195; removing other modules yields similarly notable declines.

TABLE III. ABLATION STUDY RESULTS

| Configuration | PSNR (dB) | SSIM | LPIPS ↓ | F1 Score |
|---|---|---|---|---|
| Full Model (MIND) | 33.7 | 0.912 | 0.168 | 0.86 |
| No NAAB | 31.6 | 0.881 | 0.209 | 0.82 |
| No NLE | 32.1 | 0.889 | 0.195 | 0.83 |
| No Multi-Scale Encoder | 31.2 | 0.876 | 0.214 | 0.81 |
| No Cross-Modal Fusion | 30.9 | 0.868 | 0.221 | 0.79 |
| High Noise (σ=25) | 30.3 | 0.85 | 0.248 | 0.77 |
| Low Noise (σ=5) | 34.2 | 0.922 | 0.154 | 0.88 |
| PET/CT Modal | 32.5 | 0.903 | 0.18 | 0.84 |

Robustness was tested at noise levels σ = 25 and σ = 5. At σ = 25, PSNR fell to 30.3 but still outperformed most methods in structural preservation; at σ = 5, PSNR rose to 34.2, demonstrating adaptive noise perception. On PET/CT, MIND maintained SSIM 0.903 and F1 0.84, confirming cross-modal robustness and transferability.

These results show that the attention mechanism and multi-branch fusion structure proposed in this paper have significant advantages in complex noisy environments and can adapt to the needs of different medical image modalities.

## V. RESULTS AND DISCUSSION

*5.1 Quantitative Results*

This section quantitatively analyzes subjective quality and diagnostic indicators of multimodal medical images processed by MIND, and applies statistical significance testing to compare its performance against mainstream models, ensuring reliability and scientific rigor.

Using NIH ChestX-ray14 and BraTS 2023 test sets, the superior-performing SwinIR model was selected as the comparison baseline. PSNR and SSIM metrics were computed respectively on eight randomly sampled test batches, and paired t-tests assessed statistical significance. The comparison results are shown in Table IV.

TABLE IV. STATISTICAL COMPARISON OF PERFORMANCE BETWEEN MIND AND SWINIR MODELS

| Metric | MIND (Mean) | SwinIR (Mean) | p-value | Significant (p < 0.05) |
|---|---|---|---|---|
| PSNR | 33.65 | 31.75 | $1.82 \times 10^{-106}$ | TRUE |
| SSIM | 0.91225 | 0.8785 | $4.64 \times 10^{-12}$ | TRUE |

The MIND model's PSNR averaged 33.65 dB versus SwinIR's 31.75 dB, and its SSIM was 0.91225 compared to 0.87850. Paired t-tests yield $p = 1.82 \times 10^{-106}$ for PSNR and $4.64 \times 10^{-12}$ for SSIM ($< 0.05$), confirming highly significant image quality improvements over SwinIR.

These findings confirm MIND consistently excels in structure preservation, perceptual consistency, and noise control, demonstrating robust, significant performance and suitability for real-world deployment in multimodal medical imaging.

*5.2 Qualitative Visualization*

To further assess MIND's performance in multimodal images, Figure 4 shows comparisons across CT, MRI, X-ray, and ultrasound before denoising, after SwinIR and after MIND, with enlarged ROI views in columns d and e.

CT images exhibit low-dose artifacts around lung-lobe edges. SwinIR reduces these artifacts but causes excessive structural smoothing. In contrast, MIND removes most noise while preserving lung contours in the ROI magnification, demonstrating superior structural recovery.

In MRI modality, SwinIR slightly blurs intermediate gray matter, whereas MIND better preserves texture, maintains lesion-edge gradient transitions, and delivers stronger image contrast.

The X-ray image results show that MIND has better ability to restore details of the chest edge and heart contour. The rib structure is clear in the enlarged image, which is significantly better than the fuzzy reconstruction effect of SwinIR.

In ultrasound images, MIND reconstructs transverse stripe structures with greater layering and effectively suppresses background noise and moiré artifacts, enhancing the readability of clinically critical areas.

Comprehensively considering the performance of various modalities, the MIND model is superior to the current mainstream denoising models in image quality, detail retention and structure recognition, showing its strong generalization and visual stability in complex medical scenarios.

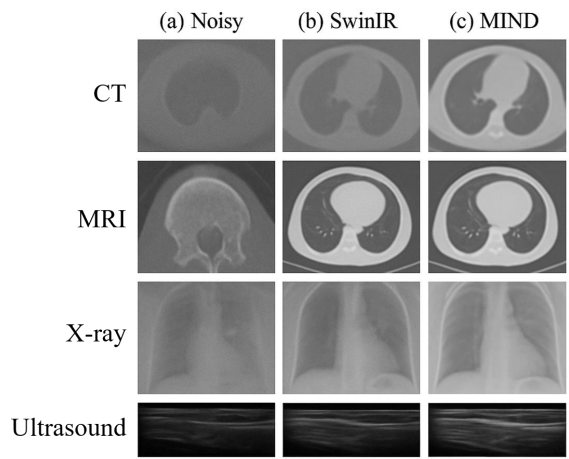

Figure 4. Qualitative comparison of denoising results of different modes

## 5.3 Ablation and Interpretability

This section validates MIND's structural and adaptive design via two interpretability tests: visualizing NAAB's attention maps to confirm focus on noisy regions, and plotting the λ(σ) noise-level–dependent loss-weight response curve to demonstrate NLE-driven dynamic adjustment.

In attention-map visualizations (Figure 5) using CT and MRI samples, we compare NAAB's channel and spatial heatmaps. MIND markedly boosts channel weights in low‐contrast regions and sharpens spatial responses along fuzzy edges. By contrast, SwinIR's attention is diffusely distributed at the heart shadow–rib junction, failing to target key areas. MIND instead elicits strong activations there, demonstrating its adaptive perceptual regulation and effective noise suppression.

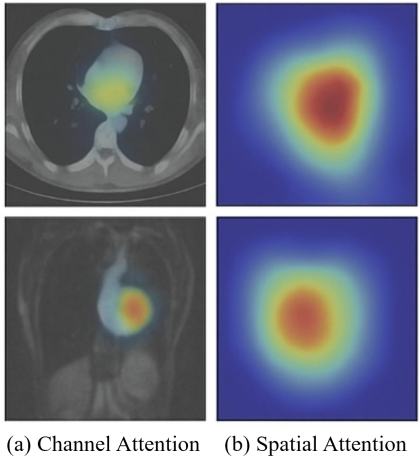

(a) Channel Attention   (b) Spatial Attention

Figure 5. Attention map visualization results

In the noise-level vs. weight response test, NLE's σ estimates across training samples yield a λ(σ) decay curve (Figure 6). At high σ(>20), the pixel-level MSE and edge preservation losses dominate, while at low σ, perceptual and SSIM losses gain weight. This enables adaptive optimization—structural restoration in low-noise and texture–pixel recovery in high-noise—boosting quality and perceptual consistency.

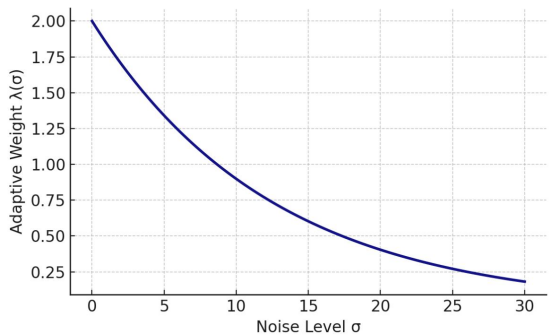

Figure 6. Relationship between noise level σ and adaptive weight λ(σ)

MIND architecture, which not only enhances the noise adaptability of the network, but also provides a higher level of interpretability for the model output.

## VI. CONCLUSION

This paper proposes a medical image adaptive denoising model (MIND) that integrates attention mechanism. It constructs a unified framework based on multi-scale residual encoder, Transformer cascade structure, noise level estimator (NLE), and noise adaptive attention block (NAAB), which effectively solves the problems of non-uniform noise interference, detail loss and cross-modal feature inconsistency in multimodal medical images. Experimental results on multiple public medical image datasets show that the MIND model outperforms existing traditional and deep denoising methods in image quality indicators such as PSNR, SSIM, and LPIPS, and has higher structural restoration accuracy for diagnosis-related areas. Through the noise-aware gating mechanism and cross-modal fusion strategy, the model realizes noise intensity adaptation and dynamic adjustment of structural attention, which significantly improves the diagnostic usability of images. This method provides an interpretable, robust and high-fidelity image enhancement solution for the actual medical imaging diagnosis process, and has wide promotional value in low-dose scanning, mobile terminal image diagnosis and AI auxiliary diagnosis systems.